\documentclass{article}
\usepackage{spconf,amsmath,graphicx}
\usepackage{verbatim}
\usepackage{booktabs}
\usepackage{multirow}
\usepackage{cite}
\usepackage{xcolor}
\usepackage{amsfonts}

\title{CASS-NAT: CTC alignment-based Single Step Non-autoregressive Transformer for speech recognition}
\name{Ruchao Fan$^{1,2}$\sthanks{\hspace{1pt} Work done as an intern at PAII Inc.}, Wei Chu$^1$, Peng Chang$^1$, Jing Xiao$^1$}
\address{$^1$ PAII Inc., USA \\
$^2$Department of Electrical and Computer Engineering, University of California Los Angeles, USA}
\begin{document}
\small
\maketitle
\begin{abstract}

We propose a CTC alignment-based single step non-autoregress- \\
ive transformer (CASS-NAT) for speech recognition. 
Specifically, the CTC alignment contains the information of (a) the number of tokens for decoder input, and (b) the time span of acoustics for each token. 
The information are used to extract acoustic representation for each token in parallel, referred to as token-level acoustic embedding which substitutes the word embedding in autoregressive transformer (AT) to achieve parallel generation in decoder.
During inference, an error-based alignment sampling method is proposed to be applied to the CTC output space, reducing the WER and retaining the parallelism as well. 
Experimental results show that the proposed method achieves WERs of 3.8\%/9.1\% on Librispeech test clean/other dataset without an external LM, and a CER of 5.8\% on Aishell1 Mandarin corpus, respectively\footnote{They are the best performing results for single step NAT in terms of WER to our knowledge at the time of submission.}.
Compared to the AT baseline, the CASS-NAT has a performance reduction on WER, but is 51.2x faster in terms of RTF. 
When decoding with an oracle CTC alignment, the lower bound of WER without LM reaches 2.3\% on the test-clean set, indicating the potential of the proposed method.
\end{abstract}

\section{Introduction}
\label{sec:intro}

In recent years, end-to-end (E2E) speech recognition models have shown competitive results \cite{luscher2019rwth, li2020comparison} compared to hybrid systems. Such models are attention-based encoder decoder (AED) \cite{chan2016listen, dong2018speech, zhou2019improving}, connectionist temporal classification (CTC) \cite{graves2006connectionist} and its variants \cite{rao2017exploring, sak2017recurrent, variani2020hybrid}.

The decoder in AED generates the output sequence autoregressively. To accelerate it, non-autoregressive transformer (NAT) was proposed for parallel generation of output sequence \cite{gu2018non, shu2020latent, saharia2020non}.
According to the number of iterations for parallel generation, NAT can be categorized into: (i) iterative NAT, and (ii) single step NAT. 
Prevailing iterative NATs regard the decoder as a masked language model, 
where the tokens with low confidence are first masked and then new tokens are predicted from unmasked tokens. The two steps are conducted alternately within a constant number of iterations \cite{chen2019non,higuchi2020mask,fujita2020insertion}. Length prediction of the decoder input is difficult for NAT models. To address the issue, Higuchi et al. used the length of recognition results of CTC \cite{higuchi2020mask}, while Chan et al. proposed a model named as Imputer, which directly used the length of input feature sequence \cite{chan2020imputer}. 
Compared to the left-to-right generation order in an autoregressive transformer (AT), an iterative NAT essentially adopts a different token generation order, where the iterations are still required. 
Single step NAT, however, can generate the output sequence in one iteration, where the word embedding is substituted with the acoustic representation for each output, assuming that the language semantics can be captured by the acoustic representations \cite{bai2020listen, tian2020spike}. Specifically, Bai et al. used a fixed length of decoder input to extract data-driven acoustic representation for decoder input \cite{bai2020listen}. 
Tian et al. used only the encoder output with CTC spikes as decoder input \cite{tian2020spike}.

Instead of using a data-driven or incomplete acoustic representation that has no physical meaning, we are trying to extract a token-level acoustic embedding, a more complete acoustic representation for each token, to capture the language semantics more effectively.
We, therefore, propose to use CTC alignment as auxiliary information for decoder to extract the token-level acoustic embedding, so as to achieve the goal of single step NAT, referred to as (CASS-NAT).
The CTC alignment is first transformed into a trigger mask to specify the time span of the acoustics for each token, as well as the length of a positional encoding. 
Then the token-level acoustic embedding for each token can be extracted in parallel by using the trigger mask and the positional encoding using encoder-decoder attention. 
During training, Viterbi-alignment is utilized as a maximum approximation of the objective function. 
During inference, an error-based alignment sampling method is proposed to improve the performance and retain the parallelism in decoder as well, where the improvements come from the possibility of sampling the candidates who may have more accurate length prediction of the decoder input.
From the analysis of the proposed sampling method, we conclude that the accurate length estimation of the decoder input could be essential for the WER reduction of NAT. As a result, the proposed CASS-NAT achieves the state-of-art NAT results on both Librispeech and Aishell1 corpus. We also derive a probabilistic framework for the proposed method.




\section{Algorithm in CASS-NAT}

The proposed CASS-NAT model architecture is based on the hybrid CTC-Attention architecture~\cite{watanabe2017hybrid} with transformer structure. The modifications are listed as below.

\subsection{Mask in attention}
\label{sec:nast}
\begin{figure}[tp]
\centering
\centerline{\includegraphics[width=0.48\textwidth]{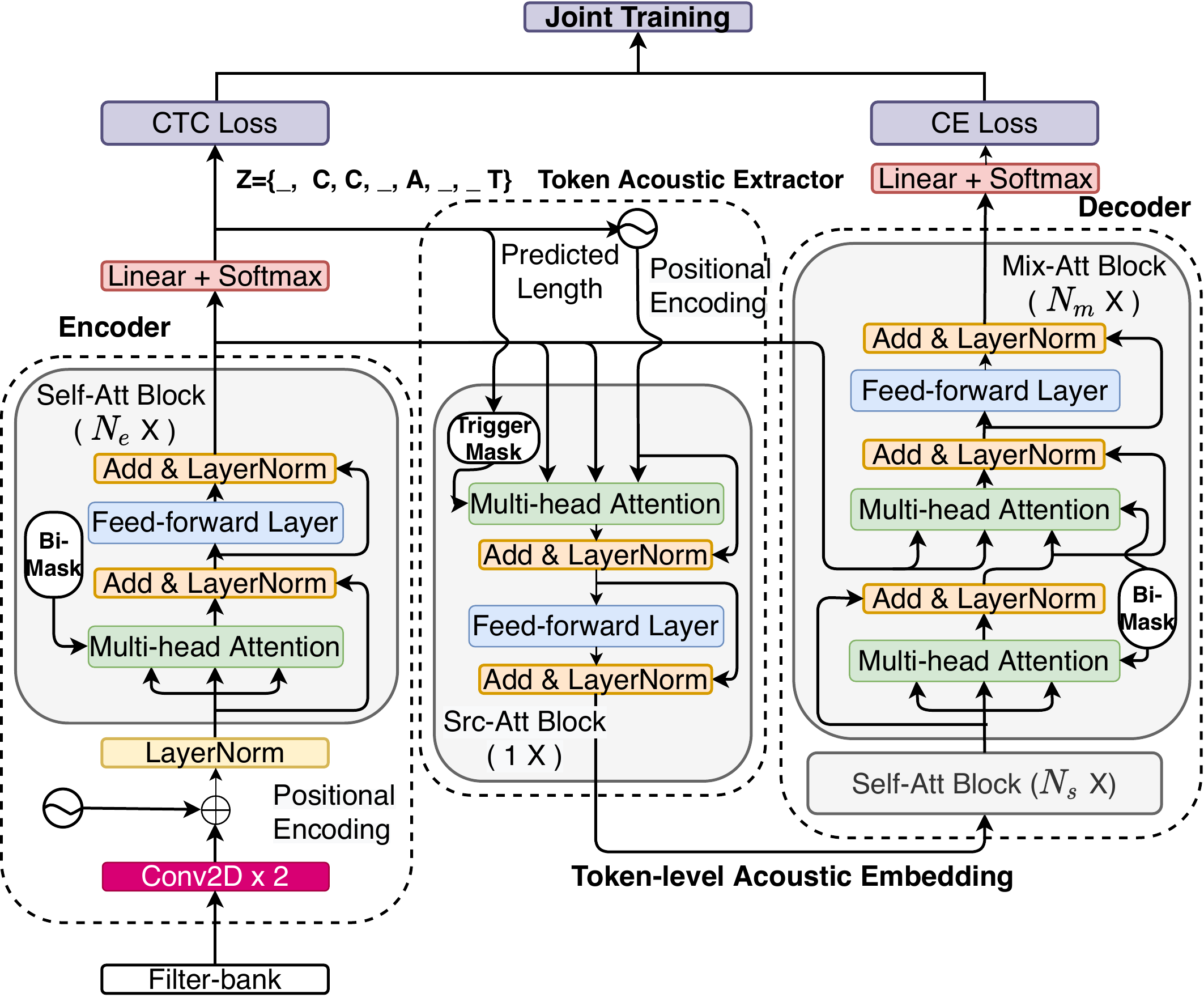}}
\caption{The proposed CASS-NAT architecture.} 
\label{fig:nast}
\end{figure}

The most basic computation in the transformer, scaled dot-product attention, is modified as:
\begin{equation}
\label{eq:attention}
    Attention(Q, K, V, M) = Softmax(\frac{QK^T}{\sqrt{d_k}})\otimes M \cdot V
\end{equation}
where $Q \in R^{n_q\times d_q}$, $K \in R^{n_k\times d_k}$, $V \in R^{n_v\times d_v}$, and $M \in R^{n_q \times n_k}$ are the queries, keys, values and mask matrix, respectively. 
The mask is used to controlled the attention range for each output position. In Fig.~\ref{fig:nast}, $M$ represents the Bi-Mask in Encoder, the Trigger Mask in Token Acoustic Extractor, and the Bi-Mask in Decoder.

\subsection{Encoder}
\label{ssec:encoder}

As shown in Fig.~\ref{fig:nast}, the encoder is the same as AT encoder \cite{dong2018speech}. On the top of the encoder, a CTC loss function is added. In this work, we refer to all the output sequence from CTC as alignment, although the alignment is commonly used for the relationship between frames and the ground truth.

\subsection{Token acoustic extractor}
\label{ssec:TAE}
The token acoustic extractor is designed to extract token-level acoustic embedding with the CTC alignment as auxiliary information.
The first information from CTC alignment is the time span of acoustics for each token. We use the information by defining a mapping from the CTC alignment to a trigger mask, where the first index of each token in the alignment is considered as its end boundary. For example, if a CTC alignment is $Z=\{\_, C, C, \_, A, \_, \_, T, \_\}$ with $\_$ as the blank symbol, the end boundary for $C$ and $A$ is $Z_2$ and $Z_5$, and thus the trigger mask for token $A$ is $[0,0,1,1,1,0,0,0,0]$. The mapping should be consistent in training and decoding.
The trigger mask in this paper is different from that used in \cite{moritz2020streaming}, where the trigger mask is to achieve streaming AT and thus the acoustic histories could be reused for each token (the trigger mask is $[1,1,1,1,1,0,0,0,0]$ for token $A$). 

The second information from CTC alignment is the number of tokens for decoder input. After removing the blank symbols and repetitions, the number of tokens in $Z$ is regarded as the predicted length of sinusoidal positional encoding, which is also the length of the decoder input. As shown in Fig.~\ref{fig:nast}, the token-level acoustic embedding for each token is then extracted with the trigger mask and the sinusoidal positional encoding using a one-layer source-attention block. Regarding the length of decoder input, the number of CTC spikes is used in \cite{tian2020spike}, resulting in an incomplete training of decoder due to the length mismatch between the CTC spikes and the ground truth. Differently, we use the viterbi-alignment in training to address the length mismatch issue.




\subsection{Decoder}
\label{ssec:decoder}
The word embedding in AT, which cannot be obtained simultaneously during decoding, is the main obstacle to achieve non-autoregressive mechanism. 
Instead, the token-level acoustic embedding has a good property of parallel generation and thus is used as a substitution of word embedding for decoder input. Since there is no need to create recurrence in decoder, unidirectional mask in the decoder is not used. We, therefore, utilize a bidirectional mask, trying to model the relations between token-level acoustic embeddings.
We assume that the token-level acoustic embedding has the same capability of capturing language semantics with word embedding since they are all token-based embeddings. In the second MHA layer of the mix-attention block, we keep the same bidirectional mask as the one used in the encoder since $K,V$ are encoder outputs.

\subsection{Training}
\label{ssec:training}
In our framework, the CTC alignments $Z$ are introduced as latent variables.
Suppose the input sequence is $X=\{x_1,...,x_t,...,x_T\}$ and the ground truth is $Y=\{y_1,...y_u,...,y_U\}$, then the objective function can be decomposed into:
\begin{equation}
\label{eq:decompose}
\begin{aligned}
 \log P(Y|X) = \log \mathbb{E}_{Z|X}[P(Y|Z,X)],  \quad Z \in q.
\end{aligned}
\end{equation}
where $q$ is the set of alignments which can be mapped to $Y$, a subset of all the alignments from CTC output space. To reduce computational cost, the maximum approximation \cite{zeyer2020new} is applied:
\begin{equation}
\label{eq:maxapprox}
\begin{aligned}
   \log P(Y|X) &\geq \mathbb{E}_{Z|X}[\log P(Y|Z,X)] \\
   &\approx \max_Z{\log\prod_{u=1}^U{P(y_u|z_{t_{u-1}+1:t_{u}}, x_{1:T})}}
\end{aligned}
\end{equation}

where $t_{u}$ is the end boundary of token $u$ described in Section \ref{ssec:TAE} and $t_0=0$. The alignment $Z$ tells the decoder where to extract the acoustic representation for each token as auxiliary information. 
The capability of language modelling for the framework is described in Section \ref{ssec:decoder}. And the same assumption as in \cite{bai2020listen} is held that acoustic embedding can also be used to capture the language semantics.

The framework is then trained by jointly maximizing the decoder loss in Eq. \ref{eq:maxapprox} and the CTC loss on the encoder side with a task ratio $\lambda$ \cite{watanabe2017hybrid} and $L_{\text{dec}}=\max_Z{\log \prod_{u=1}^U{P(y_u|z_{t_{u-1}+1:t_{u}}, X)}}$ as:
\begin{equation}
    L_{\text{joint}} = L_{\text{dec}} + \lambda \cdot \log \sum_{Z\in q}{\prod_{i=1}^T{P(z_i|X)}}
\end{equation}

\subsection{Inference: Error-based Sampling Decoding}
\label{ssec:esa_decoding}
During decoding, it is essential to obtain a CTC alignment that is close to hypothetical Viterbi-alignment. The hypothetical Viterbi-alignment is the most probable CTC alignment in Viterbi decoding assuming the transcription is available, referred to as oracle alignment in the experimental results.  
Three different approaches are proposed for alignments generation. 
The first two approaches are borrowed from the experience of CTC decoding. One is best path decoding which retains the tokens with the biggest probability of each frame, referred to as best path alignment (BPA) in this work. The other is beam search decoding over CTC output space, referred to as beam searched alignment (BSA).
Compared with BPA, BSA is supposed to generate an alignment that is closer to the oracle alignment, which could lead to a lower WER, but the parallelism would be destroyed, resulting in a significant increase of RTF.
Considering the expectation in Eq. \ref{eq:maxapprox}, we use a third alignment generation approach by sampling from CTC output space. Compared to BPA, the sampled alignments have the possibility of being closer to the oracle alignment. Compared to BSA, the sampling can be implemented in parallel, avoiding the increase of RTF. Finally, either the AT baseline or the language models can be used to score and identify the best overall alignment.
%
Sampling from CTC output is not easy due to the large sampling space. However, we observe that 
BPA tends to make errors in the CTC output where the Top-1 probability is low.
We therefore propose an error-based alignment sampling method, which is more effective in terms of WER and RTF compared to the BPA and BSA, referred to as ESA.


\begin{figure}[tp]
\centering
\centerline{\includegraphics[width=0.48\textwidth]{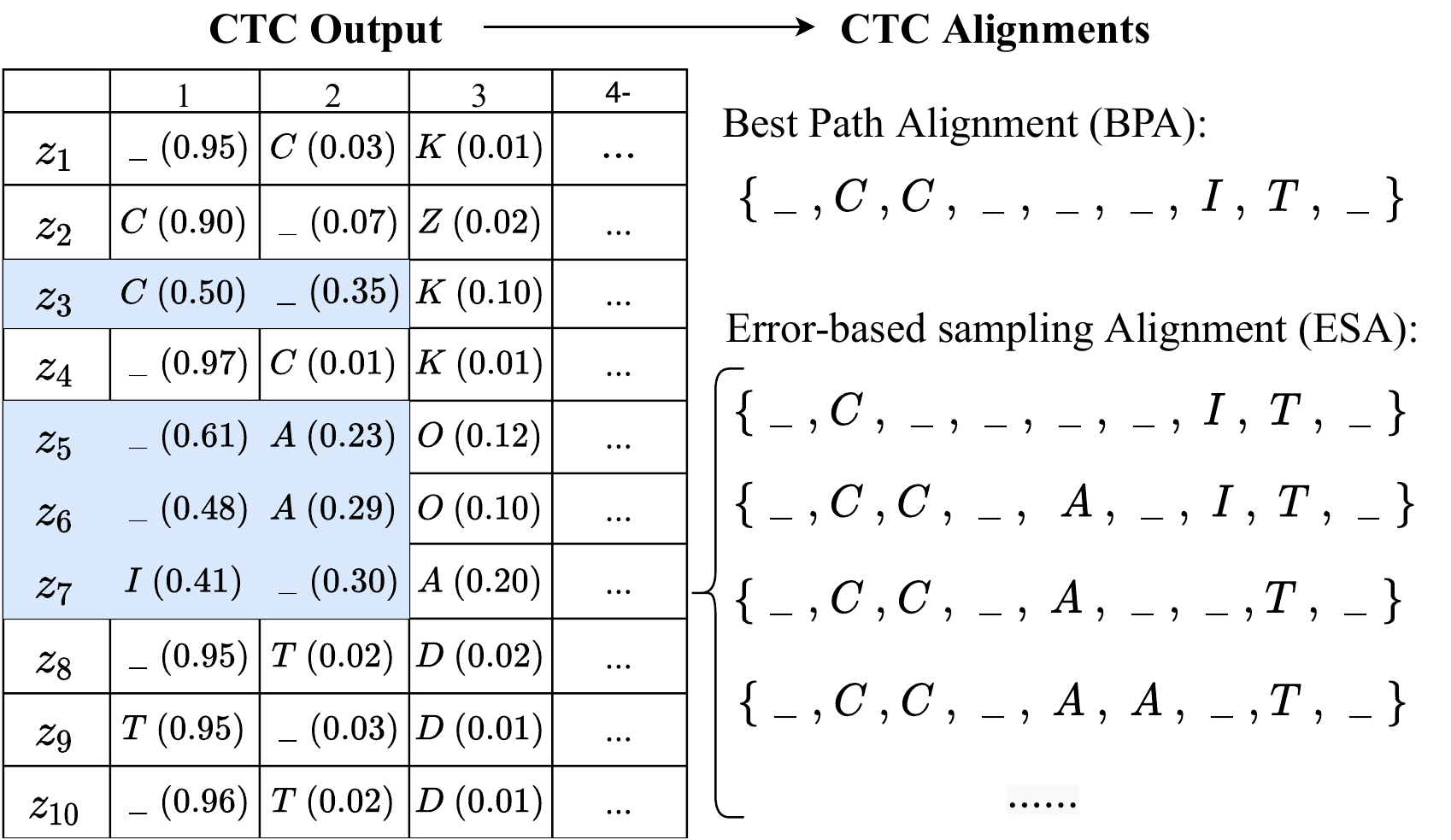}}
\caption{Illustration of error-based alignment sampling method. $C(0.90)$ indicates $P(z_i=C|X)=0.90$.}
\label{fig:esa_decode}
\end{figure}

To generate ESA, we first select the CTC outputs with low Top-1 probability (the threshold is empirically set to 0.7), shown as blue part in Fig.~\ref{fig:esa_decode}. Then the Tope-2 tokens are randomly chosen in each selected output while other outputs keep using the Top-1 token. For instance, the CTC output $z_3$, $z_5$, $z_6$, $z_7$ are selected, and 4 example alignments are sampled as shown in Fig. \ref{fig:esa_decode}. The reason for sampling within the Top-2 tokens is that the trigger mask of each token is only sensitive to whether the token in the alignment is blank or not, and we observe that most errors in BPA contain blank in Top-2 tokens. ESA significantly reduces the sampling space and aims at correcting the output that is prone to make errors. The sampling in decoding will not affect much of the inference speed because all the alignments can be generated in parallel to compute the decoder input.
Note that ESA can have different length for decoder input compared to BPA. As shown in Fig.~\ref{fig:esa_decode}, the length is 4 for BPA, while the lengths are 3, 5, 4, and 5 in the case of ESA. This fluctuation of the token number in the alignment allows the ESA method to possibly sample an alignment that is of the same length as the oracle alignment.


\section{Experiments}
\label{sec:expsetup}

\subsection{Experimental setup}
\label{ssec:setup}

The experiments were conducted on the 960-hour LibriSpeech English corpus \cite{panayotov2015librispeech} and the 178-hour Aishell1 Mandarin corpus \cite{bu2017aishell}. All experiments used 80-dim Mel-filter bank features, computed every 10ms with a 25ms window. Every 3 consecutive frames were concatenated to form a 240-dim feature vector as the input. The sets of output labels consist of 5k word-pieces obtained by the SentencePiece method \cite{kudo2018sentencepiece} for Librispeech and 4230 Chinese characters for Aishell1 which was obtained from the training set. 

A CTC/Attention AT baseline was first trained with the same architecture ($N_e=12$, $N_d=6$, $d_{FF}=2048$, $H=8$, $d_{MHA}=512$) as in \cite{karita2019comparative} for Librispeech. For CASS-NAT, the decoder in AT was replaced by $N_{s}=3$ self-attention blocks and $N_m=4$ mix-attention blocks. For Aishell1, both the AT baseline and CASS-NAT have the same architecture as their counterparts used for Librispeech except $N_e=6$.  
For all methods, each of the two CNN in the encoder has 64 filters with kernel size of $3\time 3$ and stride of 2, leading to a 4x frame rate reduction.
For all models, the Double schedule in \cite{park2019specaugment} was adopted as the learning schedule, where the learning rate is ramped up to and held at 0.001, then be exponentially decayed to 1e-5. Layer normalization, dropout with rate of 0.1 and label smoothing with a penalty of 0.1 were all applied as the common strategies for training a transformer. We also applied spec-augment \cite{park2019specaugment} for fair comparisons with the results in literature. We additionally applied speed-perturbation for Aishell1. Encoder initialization was used for CASS-NAT training. The task ratio $\lambda$ in Section \ref{ssec:training} was set to 1 for all models. We used development sets for early stopping and model averaging for final evaluation. Most of the experiments ended within 90 epochs. 
The evaluation of the inference speed was conducted on an NVIDIA Tesla V100 GPU with the batch size of one.

Additionally, a transformer-based language model was trained with the provided text in Librispeech. Both the settings of the model and ctc-attention decoding were the same as in ESPNet \cite{watanabe2018espnet}. For NAT decoding with the language model, the beam size was set to 5.


\subsection{Accuracy and speed analysis}
\label{ssec:results}

\begin{table}[t]
\caption{\small {A comparison of accuracy and speed of Autoregressive Transformer (AT) and non-AT (NAT) algorithms on Librispeech. \textbf{BPA}: decoding with the best path alignment. \textbf{BSA}: beam search alignment. \textbf{ESA}: the proposed error-based sampling alignment in which 50 alignments are sampled. \textbf{RTF}: Real-time factor. The LM used in CASS-NAT and AT are the same but trained with fewer epochs than the LM used in ESPNet.}}
\scriptsize
\centering
\begin{tabular}{c|c| c cccc c}
\hline
\multicolumn{2}{c}{\multirow{3}{*}{}} & \multirow{3}{*}{Type} & \multicolumn{4}{c}{WER (\%)} & RTF \\
\cmidrule(r){4-7} \cmidrule(r){8-8}
\multicolumn{2}{c}{} & ~ & dev- &  dev- &  test- &  test- & \ test- \\
\multicolumn{2}{c}{} & ~ & clean &  other &  clean &  other & clean \\
\hline \hline
\multicolumn{7}{l}{\textbf{Without LM}} & \multicolumn{1}{c}{} \\
\hline
\multicolumn{2}{c}{RETURNN \cite{luscher2019rwth}} & AT & 4.3 & 12.9 & 4.4 & 13.5 & - \\
\multicolumn{2}{c}{ESPNet} & AT & 3.2 & 8.5 & 3.6 & 8.4 & - \\
\multicolumn{2}{c}{AT (ours)} & AT & 3.4 & 8.5 & 3.6 & 8.5 & 0.562 \\
\multicolumn{2}{c}{Imputer \cite{chan2020imputer}} & NAT & - & - & 4.0 & 11.1 & - \\
\hline
\multirow{3}{*}{CASS-NAT} & BPA & NAT & 4.4 & 10.6 & 4.5 & 10.7 & 0.005 \\
~ & BSA & NAT & 3.9 & 9.6 & 3.9 & 9.6 & 0.655 \\
~ & ESA & NAT & 3.7 & 9.2 & 3.8 & 9.1 & 0.011 \\
\hline
\multicolumn{7}{l}{\textbf{With LM}} & \multicolumn{1}{c}{} \\
\hline
\multicolumn{2}{c}{RETURNN \cite{luscher2019rwth}} & AT & 2.6 & 8.4 & 2.8 & 9.3 & - \\
\multicolumn{2}{c}{ESPNet \cite{karita2019comparative}} & AT & 2.3 & 5.6 & 2.6 & 5.7 & - \\
\multicolumn{2}{c}{AT (ours)} & AT & 2.5 & 5.7 & 2.7 & 5.8 & - \\
\hline
CASS-NAT & ESA & NAT & 3.3 & 8.0 & 3.3 & 8.1 & - \\
 \hline
\end{tabular}
\label{tab:results1}
\end{table}

\begin{table}[t]
\caption{\small A comparison of WERs on Aishell1 with the existing works.}
\small
\centering
\begin{tabular}{c c c c}
\hline
CER(\%) & NAT Type & Dev & Test \\
\hline \hline
AT (ours) & n/a & 5.5 & 5.9 \\
Masked-NAT \cite{chen2019non} & iterative & 6.4 & 7.1 \\
Insertion-NAT \cite{fujita2020insertion} & iterative & 6.1 & 6.7 \\
ST-NAT \cite{tian2020spike} & single step & 6.9 & 7.7 \\
LASO \cite{bai2020listen} & single step & 5.8 & 6.4 \\
CASS-NAT (ours) & single step & 5.3 & 5.8 \\
\hline
\end{tabular}
\label{tab:results2}
\end{table}

Table \ref{tab:results1} presents the results of AT baseline and Non-AT models on the Librispeech dataset\footnote{Espnet just updated the AT structure to Conformer \cite{gulati2020conformer}. We would like to update the CASS-NAT structure as well in the future work.}.
First, for the proposed CASS-NAT method, the ESA decoding reduced WER significantly compared to both BPA and BSA, although it had a moderate increase of RTF over BPA due to the ranking process for multiple sampled alignments. Thus, we use the results of ESA for CASS-NAT by default unless otherwise stated. Then in the case of no external LM, compared to the AT baseline, the WER of the proposed CASS-NAT was 6\% higher relatively, but was 51.2x faster in terms of RTF, which is expected. Compared to the iterative NAT-based Imputer, the CASS-NAT achieved a better performance on WER, and theoretically was faster than the Imputer, which needs 8 iterations during decoding. When decoding with an external LM, the gap of WER between AT baselines and the proposed CASS-NAT was increasing. The reason could be that AT benefited more from the external LM due to the same autoregressive mechanisms in AT and the LM.


On Aishell1 Mandarin task, Table~\ref{tab:results2} shows that the proposed CASS-NAT had a lower WER than both our AT baseline and WERs reported in other NAT methods, which suggests the effectiveness and generalizability of the proposed method. 


%

\subsection{Analysis of the CTC Alignments}
\label{ssec:analysis}

Further, we analyzed the alignments obtained from different decoding strategies and hopefully to comprehend why ESA can improve the accuracy. 
Two metrics were evaluated on the test sets of Librispeech, mismatch rate (MR) and length prediction error rate (LPER), measured between the generated alignments and the oracle alignment. The two metrics were computed after the blank and repetitions in the alignment were removed because of the mapping rule in Section~\ref{ssec:TAE}. For MR, only deletion and insertion errors were counted as mismatch since substitution errors will not change the end boundaries and the number of predicted tokens. If the number of tokens in an alignment is different from that in the oracle alignment, this alignment is considered as a case of length prediction error.

\begin{table}[tp]
\caption{\small {A comparison of different alignment generation methods in CASS-NAT decoding without LM. \textbf{Oracle}: Viterbi-alignment with ground truth. \textbf{MR}: mismatch rate; \textbf{LPER}: length prediction error rate. \textbf{S}: the number of alignments sampled in ESA.}}
\footnotesize
\centering
\begin{tabular}{c c cc cc cc}
\hline
\multirow{2}{*}{Alignment} & \multirow{2}{*}{S} & \multicolumn{2}{c}{WER (\%)} & \multicolumn{2}{c}{MR (\%)} & \multicolumn{2}{c}{LPER (\%)} \\
\cmidrule(r){3-4} \cmidrule(r){5-6} \cmidrule(r){7-8} 
~ & ~ & test- & test- & test- & test- & test- & test- \\
~ & ~ & clean & other & clean & other & clean & other \\
\hline \hline
Oracle & n/a & 2.3 & 5.8 & n/a & n/a & n/a & n/a \\
\hline
BSA & n/a & 3.9 & 9.6 & 2.2 & 5.8  & 27.9 & 48.3 \\ 
BPA & n/a & 4.5 & 10.7 & 2.1 & 4.9 & 31.0 & 51.8 \\
\hline
\multirow{4}{*}{ESA}  & 10  & 3.9 & 9.4 & 2.9 & 5.7 & 26.4 & 42.8 \\
~  & 50  & 3.8 & 9.1 & 3.1 & 5.8 & 25.3 & 41.9 \\
~  & 100 & 3.8 & 9.0 & 3.0 & 5.8 & 25.1 & 41.8 \\
~  & 300 & 3.8 & 9.0 & 3.1 & 5.8 & 25.1 & 41.9 \\
\hline
\end{tabular}
\label{tab:analysis}
\end{table}

\begin{figure}[t]
\centering
\centerline{\includegraphics[width=0.4\textwidth]{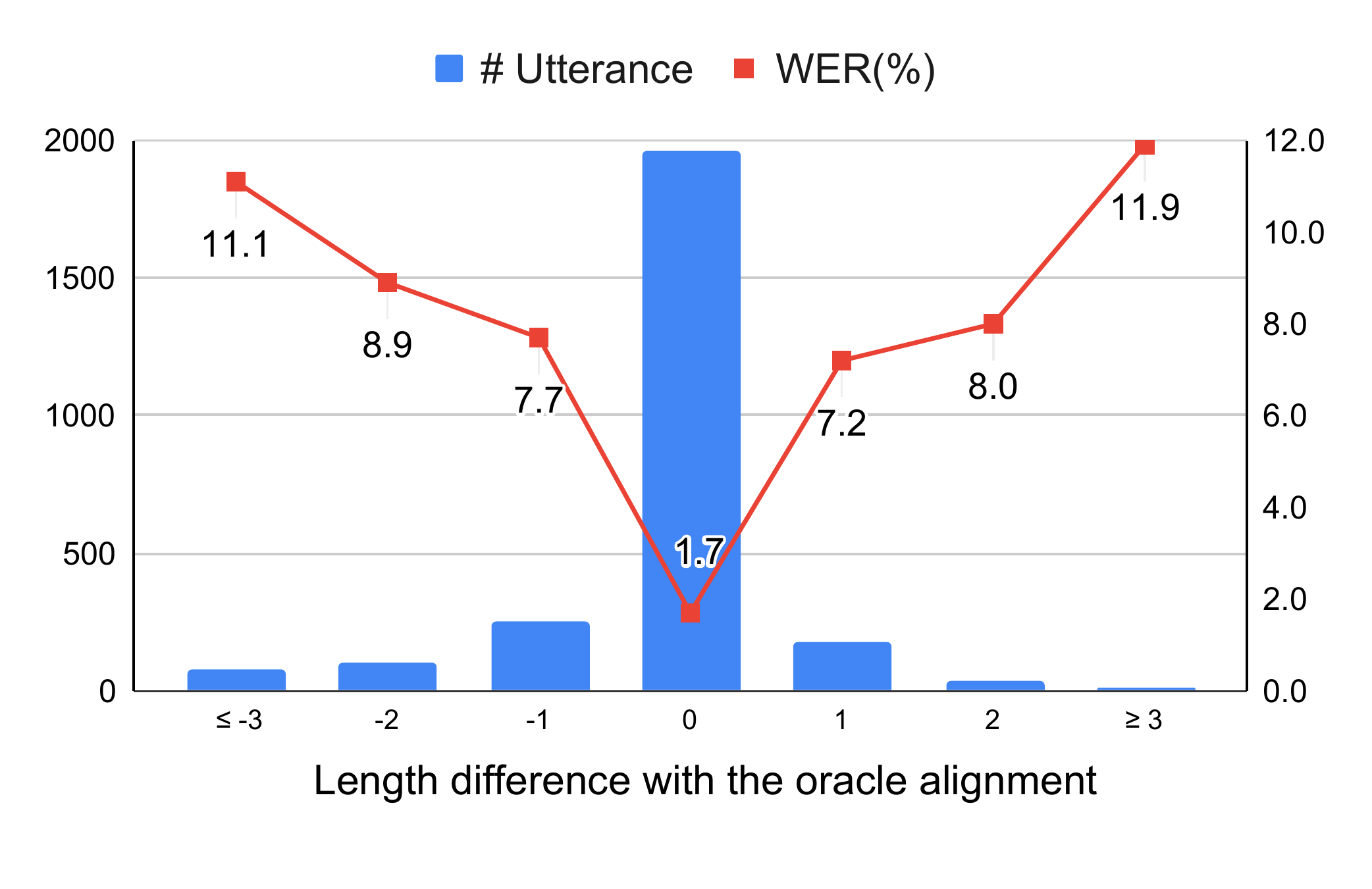}}
\caption{Length prediction error distributions and corresponding WERs with ESA(s=50) decoding on the test-clean dataset.}
\label{fig:errdist}
\end{figure}

Results are presented in Table \ref{tab:analysis}. We can see that the CASS-NAT had a lower bound with WER of 2.3\% when using the oracle alignment, showing that the framework is promising. For ESA, no further gains were observed when the number of sampled alignment is greater than 50. We can also observe from the table that the WER was more correlated to the LPER other than MR. It suggests that a correct estimation of the decoder input length is more important for NAT, which is also mentioned in \cite{chen2019non}. Fig.~\ref{fig:errdist} again shows the importance of length prediction accuracy. We can see that the WER was lower than 2\% when the length of the decoder input was estimated correctly. The analysis suggests a possible guideline on how to find a more effective alignment sampling strategy besides ESA. It also shows the potential of CASS-NAT to get a more promising result.

\section{Conclusions}
\label{sec:conclusion}

This work presented a novel CASS-NAT framework which leverages the CTC alignment to extract the token-level acoustic embedding as a substitution of the word embedding in autoregressive models. During the training, Viterbi-alignment was considered as a maximum approximation of the posterior probability. During the inference, we proposed an error-based alignment sampling method to reduce the mismatch between the generated alignment and the oracle alignment. We also showed the importance of the prediction for the length of decoder input in NAT through the analysis of the alignments. As a result, the proposed CASS-NAT had $\sim$50x faster inference speed than autoregressive transformer. To the best of our knowledge, the results of CASS-NAT are the best-reported ones for single step NAT methods so far on both the Librispeech (WERs of 3.8\%/9.1\%) and Aishell1 (a CER of 5.8\%) when no external LM is used.
\bibliographystyle{IEEEbib}
\bibliography{refs}

\end{document}